\documentclass[aip, amsmath, amssymb, superscriptaddress, reprint]{revtex4-1}
\usepackage{graphicx} %include figure files
\usepackage{dcolumn} %align table columns on decimal point (?)
\usepackage{bm} %bold math
\usepackage{placeins} % for floating point stuff
\usepackage{hyperref} %add hypertext capabilities
\usepackage{hhline} % for double horizontal line in tables
\setcitestyle{super} % for superscript citations
\usepackage{color} % for text coloring
\usepackage{comment}
\usepackage[margin=1.0in]{geometry}
\usepackage{wrapfig} % to wrap text around pictures

\begin{document}

\title{Crystallinity characterization of white matter in the human brain}

\author{Erin G. Teich}
\affiliation{Department of Bioengineering, School of Engineering \& Applied Science, University of Pennsylvania, Philadelphia PA 19104}
\author{Matthew Cieslak}
\affiliation{Department of Psychiatry, Perelman School of Medicine,  University of Pennsylvania, Philadelphia PA 19104}
\author{Barry Giesbrecht}
\affiliation{Department of Psychological \& Brain Sciences, University of California, Santa Barbara CA 93117}
\author{Jean M. Vettel}
\affiliation{Army Research Laboratory, Aberdeen Proving Ground MD 21005}
\author{Scott T. Grafton}
\affiliation{Department of Psychological \& Brain Sciences, University of California, Santa Barbara CA 93117}
\author{Theodore D. Satterthwaite}
\affiliation{Department of Psychiatry, Perelman School of Medicine,  University of Pennsylvania, Philadelphia PA 19104}
\author{Danielle S. Bassett}
\affiliation{Department of Bioengineering, School of Engineering \& Applied Science, University of Pennsylvania, Philadelphia PA 19104}
\affiliation{Department of Physics and Astronomy, College of Arts \& Sciences, University of Pennsylvania, Philadelphia PA 19104}
\affiliation{Department of Electrical and Systems Engineering, School of Engineering \& Applied Science, University of Pennsylvania, Philadelphia PA 19104}
\affiliation{Department of Neurology, Perelman School of Medicine,  University of Pennsylvania, Philadelphia PA 19104}
\affiliation{Department of Psychiatry, Perelman School of Medicine,  University of Pennsylvania, Philadelphia PA 19104}
\affiliation{Santa Fe Institute, Santa Fe NM 87501}

\date{\today}

\begin{abstract}
White matter microstructure underpins cognition and function in the human brain through the facilitation of neuronal communication.
The non-invasive characterization of this white matter structure remains a research frontier in the neuroscience community, as the advent of accurate and detailed measurements of white matter in living brains would enable better detection of disease and contribute to the growing body of knowledge regarding the influence of development and aging on brain structure.
Efforts to assess white matter microstructure, however, have been hampered by the sheer amount of information needed to fully characterize this architecture.
Current techniques within neuroimaging deal with this high dimensionality problem by representing millimieter-scale features of white matter with single scalars that are often not easy to interpret and do not contain information regarding structures on larger length scales.
Here, we address these issues by introducing tools from materials science for the characterization of white matter microstructure.
We investigate structure on a mesoscopic scale by analyzing its homogeneity and determining which regions of the brain are structurally homogeneous, or ``crystalline" in the context of materials science, versus structurally heterogeneous, or ``disordered." 
We find that structural homogeneity, or ``crystallinity," provides novel information regarding white matter that is not captured by other traditional diffusion metrics, and varies across the brain along interpretable lines of anatomical difference, with highest homogeneity in regions adjacent to the corpus callosum. 
Furthermore, crystallinity is highly reliable across multiple scans of the same individual, and yet also varies appreciably between individuals, making it a potentially useful biomarker to examine individual differences in white matter along the dimensions of sex, genetics, and physiology. 
Given these positive features, we also use structural homogeneity to parcellate white matter into ``crystal grains," or spatially contiguous sets of voxels of high structural similarity, and find overlap with other commonly used white matter parcellations. 
Finally, we characterize the shapes and rotational symmetries of local white matter diffusion signatures through another tool from materials science -- bond-orientational order parameters that were originally developed to analyze supercooled liquids and glasses-- and locate fiber crossings and white matter fascicles with this new technique.
Our results provide new means of assessing white matter microstructure on multiple length scales, and open new avenues of future inquiry.
We hope that this work fosters future productive dialogue between the fields of soft matter and neuroscience.
\end{abstract}

\maketitle

\section{Introduction}
The fields of soft matter and diffusion MRI, each relative newcomers in their respective scientific corners, may seem at first glance quite unrelated. 
Soft matter physics is grounded in statistical mechanics and typically deals in spontaneously self-organizing ensembles of particles, molecules, or other micro-scale constituents.
Diffusion MRI, by contrast, lies at the intersection of engineering and neuroscience, and is focused primarily on characterizing microstructure in a very specific macroscale organ, the brain \cite{Basser2014}.

Both fields, however, are chiefly concerned with understanding \emph{structure}, and its relationship to function.
In soft materials, local configurations of particles often determine macroscopic properties such as bulk or shear moduli \cite{Ashcroft1976, Schlegel2016}, or even the propensity of the material as a whole to crystallize upon cooling or remain in an amorphous, arrested state \cite{Royall2015}.
In diffusion imaging, the orientation of myelinated white matter structures is estimated in millimeter-scale regions of the brain based on observed magnetic resonance (MR) signal changes related to water diffusion. 
Local white matter orientation distribution functions (ODFs) can be used to track large white matter fascicles \cite{Behrens2013} between cortical regions.
White matter connectivity underlies normal brain function, and disruption of these connections is hypothesized to contribute to a variety of pathologies that may arise in abnormal development and following injury \cite{Bodini2013}.

This paper provides a link between these two communities, grounded in their common interest in local structure and consequent macroscopic behavior. 
We calculate the structural homogeneity (a metric commonly used in soft matter to characterize particle ensembles) of white matter in the human brain. Our approach complements prior efforts to determine biological complexity from local measurements \cite{Hallgrimsson2018}, and differs from those efforts by utilizing techniques explicitly grounded in materials science to understand brain organization.
We determine which regions of the brain are structurally homogeneous, or ``crystalline" in the context of materials science, and which are structurally heterogeneous, or ``disordered."
%% 1. 
We find that white matter homogeneity, or crystallinity, provides information about neuronal architecture that is independent of other structural measurements often used in the neuroimaging community such as fractional anisotropy and mean diffusivity.
%% 2.
White matter crystallinity is also reliable and reproducible across multiple imaging scans of the same subject, but variable across subjects. 
Thus, this novel metric may be a useful marker to distinguish individual differences in white matter microstructure.
%% 3.
In general, crystallinity varies throughout the brain, and is highest for its most internal white matter regions, including the corona radiata, internal capsule, corpus callosum, and uncinate fasciculus.
%% 4.
Additionally, we find that parcellation of white matter into ``crystal grains," or spatially contiguous sets of voxels of high structural similarity, results in a new white matter atlas that has partial overlap with more commonly used white matter atlases, but incorporates the distinct local morphologies of brain tissue.
Taken together, our results illustrate that crystallinity is a fruitful new means of analyzing the structure of white matter, with multiple future promising directions.

Crystallinity measures structural homogeneity of white matter in local regions throughout the brain; however, characterization of white matter on an even smaller scale (that of shape features of individual white matter elements) is a complementary and active research thrust of diffusion imaging.
Efforts to pinpoint millimeter-scale white matter structural features such as fiber crossings and coherent fascicles in diffusion imaging typically involve methods designed to capture overall diffusion signal strength and anisotropy.
We demonstrate that in this arena, as well, tools from soft matter are valuable for analysis of the overall shape of the diffusion signal, beyond its strength or anisotropy.
We characterize the shapes of local white matter diffusion signatures using bond-orientational order parameters originally developed to characterize symmetries of particle clusters in supercooled liquids and glasses \cite{Steinhardt1983}.
When applied in a neuroimaging context, these order parameters characterize local white matter symmetries and provide useful, rotationally-invariant information regarding where white matter tracts are especially aligned or where they cross.
Our results provide further evidence that approaches grounded in materials science are useful for the structural characterization of the human brain as complex soft matter on multiple length scales.
We look forward to future interdisciplinary extensions of this work, bridging soft matter and neuroimaging.

\section{Methods}

\subsection{Diffusion image analysis}

We use two datasets throughout this paper that each have particular strengths that are relevant to our distinct goals.
The first, used to illustrate our methods, is the publicly available example of a high angular resolution diffusion image (the "Stanford HARDI" dataset) with 160 gradient directions, distributed with \emph{dipy} \cite{Garyfallidis2014}.
We estimated fiber orientation distributions (FODs) on the Stanford HARDI data using constrained spherical deconvolution \cite{Tournier2007} and extract FOD peaks using \emph{dipy}.
An example slice of these FODs in an array of neighboring voxels is shown in Fig. \ref{fig:methods}A.

We chose a second data set to showcase the performance of our methods in detecting individual differences reliably across scans. Specifically, we use an extensive set of diffusion images collected as part of the Cognitive Resilience and Sleep History (CRASH) study \cite{Thurman2018}.
In addition to a host of other metrics, the CRASH study gathered 8 diffusion MRI scans for 30 subjects, taken bi-weekly. 
Subjects (13 males, 17 females) ranged in age between 18-35 years, with a mean age of 23 years.
The diffusion acquisitions used a bipolar pulse sequence to sample a Cartesian grid in $q$-space at 258 coordinates with a maximum $b$-value of 5000 s/mm$^2$ with 1.8mm isotropic voxel size. Images were preprocessed using QSIPrep \citep{Cieslak2020}, which included MP-PCA denoising \citep{veraart2016} and head motion correction. Preprocessed images were reconstructed using generalized $q$-sampling imaging (GQI) \citep{Yeh2010b}. GQI ODFs and peaks were calculated in DSI studio, and a group threshold of 0.02 was used to threshold ODF peaks. T1-weighted anatomical scans were registered to the MNI 2009 asymmetric nonlinear template using ANTs \citep{ants}.

\subsection{Crystallinity}
We characterize diffusion signal homogeneity across neighboring voxels by characterizing the ``crystallinity" of the ODF signatures.
Crystallinity measures the similarity between the structural environment of each voxel (consisting of the peak vectors extracted from each ODF signal) and the environments of its neighboring voxels. 
Voxels of especially high crystallinity contain ODF signals that are especially similar, on average, to the ODF signals in neighboring voxels.
Voxels of especially low crystallinity contain ODF signals that are dissimilar from ODF signals of neighboring voxels; these voxels are ``disordered" in the language used by the soft matter community.
To quantify crystallinity, we measure the average deviation between the diffusion signal of each voxel and those of its neighbors using an environment-matching module in the open-source analysis toolkit \emph{freud} \cite{Ramasubramani2020}.
We then normalize this deviation by the overall diffusion signal strength in the appropriate voxel, creating a dimensionless parameter ($\tilde{\Delta}$) that acts as a proxy for crystallinity.

We now provide the mathematical details of this process.
For all neighboring voxel pairs $i$ and $j$ (defined as neighbors if the voxel boundaries share faces, edges, or vertices), we measure the root-mean-squared deviation, $\Delta_{ij}$, between their environments:

\begin{align}
\Delta_{ij} = \sqrt{\langle \delta^2_{mm'} \rangle}.
\end{align}

\noindent This calculation is shown schematically in Fig. \ref{fig:methods}B.
In the above expression, $\delta_{mm'} \equiv \vert \bm{r}_{im} - \bm{r}_{jm'} \vert$ is the magnitude of the vector difference between voxel $i$'s $m$-th ODF peak vector $\bm{r}_{im}$ and voxel $j$'s $m'$-th ODF peak vector $\bm{r}_{jm'}$. 
The average $\langle \delta_{m,m'} \rangle$ is taken over the mapping $(m,m')$ found that best minimizes $\Delta_{ij}$.
To obtain this mapping, we consider each vector in the set $\{ \bm{r}_{jm'} \}$ in turn, and greedily pair it with the closest vector in the set $\{ \bm{r}_{im} \}$ that is unpaired.
If voxels $i$ and $j$ do not have equivalent numbers of ODF peak vectors, then we first augment the smaller set of vectors with $\bm{0}$ vectors until both sets are the same size before finding the optimal $(m,m')$ mapping.

From the set of root-mean-squared deviations $\Delta_{ij}$ between the environment of voxel $i$ and each of its neighbors $j$, we can define the normalized average deviation of voxel $i$'s environment from those of its neighbors:

\begin{align}
    \tilde{\Delta}_i = \frac{\langle \Delta_{ij} \rangle}{\langle r_{im} \rangle}.
\end{align}

\noindent This measure is our proxy for crystallinity for each voxel. The numerator is the average $\Delta_{ij}$ between voxel $i$ and its $N_i$ neighbors, $\langle \Delta_{ij} \rangle \equiv \frac{1}{N_i} \sum_j \Delta_{ij}$. The denominator is the average magnitude of the $M_i$ ODF peaks in voxel $i$, $\langle r_{im} \rangle = \frac{1}{M_i} \sum_m r_{im}$, excluding any augmented $\bm{0}$ vectors.
The parameter $\tilde{\Delta}_i$ is thus normalized such that it gives the average deviation of the diffusion signal $i$ from its neighboring signals as a fraction of that signal's average magnitude.
This measure is low when crystallinity is high, and \emph{vice versa}.

To gain an intuition for the spatial distribution of the measure $\tilde{\Delta}$, we show this value for each voxel in a slice through the example HARDI image described previously (Fig. \ref{fig:methods}C). 
Each voxel $i$ is colored according to $\tilde{\Delta}_i$ such that those with lower values of $\tilde{\Delta}$ (i.e., more crystalline) are lighter.
Note that white matter structure can be clearly visualized by this metric. Other interesting structural features emerge as well, such as voxels that mark ``grain boundaries" between spatially contiguous sets of voxels of high structural similarity (or ``crystal grains").

\subsection{Crystal grain parcellation}
We also use structural similarity between neighboring voxels to construct white matter parcellations that reflect homogeneous regions of brain tissue.
We refer to these structurally similar regions as crystal grains, and formally identify them as strongly intra-connected submodules, or communities, in a network representation of each brain scan.
In this representation, each voxel is a node of the network, and edges between nodes are weighted according to the structural similarity of the corresponding voxels.
Edges only exist between node pairs that correspond to neighboring voxel pairs.
Specifically, the edge weight $W_{ij}$ between nodes $i$ and $j$ is defined as:

\begin{align}
    W_{ij} = \frac{1}{\frac{\Delta_{ij}}{N_{ij}} + 1}.
\end{align}

\noindent The parameter $\Delta_{ij}$ is the root-mean-squared deviation between the diffusion signals of voxels $i$ and $j$ as defined previously. We note that we set $\Delta_{ij} = \Delta_{ji}$ for computational efficiency when computing this quantity over all neighbor pairs, and this equality always holds if $\Delta_{ij}$ is a true global minimum.
The parameter $N_{ij}^2 \equiv \langle \bm{r}^2_{im} + \bm{r}^2_{jm'} \rangle$, with the average taken over the mapping $\left( m, m' \right)$ found that best minimizes $\Delta_{ij}$. 
As before, $\bm{r}_{im}$ is voxel $i$'s $m$-th ODF peak vector.
%% double-check this
The parameter $N_{ij}$ is thus a normalization factor that represents the root-mean-squared deviation over all paired vectors if each pair of vectors was orthogonal.
%% mention at some point that using other powers doesn't really influence crystal grains- it just changes the gamma ranges over which those grains are found? (as in hardi slice analysis)
When diffusion signals are structurally identical, that is, $\Delta_{ij} = 0$, then edge weight is highest ($W_{ij} = 1$).
As the structural similarity between diffusion signals decreases and $\Delta_{ij}$ increases, the edge weight decreases to 0.

Within this network representation, we identify crystal grains of homogeneous structure by using a Louvain-like locally greedy algorithm \cite{Jeub2019} to identify communities of nodes (voxels) that are densely intra-connected and sparsely inter-connected \cite{Fortunato2010}.
The algorithm maximizes the so-called modularity \cite{Newman2004} $Q$ of the network, which is defined as:

\begin{align}
    Q = \sum_{ij} \left[ W_{ij} - \gamma P_{ij}\right] \delta (c_i, c_j).
\end{align}

\noindent The scalar $W_{ij}$ is the edge weight between nodes $i$ and $j$, $P_{ij}$ is the expected edge weight between those nodes in a suitably-chosen null model, $\delta (c_i, c_j)$ is a Kroenecker delta that is 1 if $i$ and $j$ are in the same community (\emph{i.e.} if community indices $c_i = c_j$) and 0 otherwise, and $\gamma$ is a free scalar parameter.
The parameter $\gamma$ controls the resolution over which communities are detected. As $\gamma$ goes to 0, $W_{ij} - \gamma P_{ij}$ becomes positive for all edge weights, and so maximizing $Q$ consists of grouping all connected nodes into the same community. In contrast, higher $\gamma$ provides a threshold for which edge weights will contribute to $Q$, and effectively filters communities such that they consist only of the most strongly-connected nodes.

Many different null models are employed in the literature, each specific to the data and scientific question of interest. Here, we use a geographical null model \cite{Bassett2015a, Papadopoulos2016} previously utilized to detect communities in spatially-embedded and locally-connected networks.
In this model, $P_{ij} = \rho A_{ij}$, where $\rho = \langle W_{ij} \rangle$ is the mean edge weight of the network and $A_{ij}$ is 1 if nodes $i$ and $j$ have an edge between them and 0 otherwise.
This null model effectively encodes the physical constraints experienced by the system: namely, that each voxel is only structurally compared with its spatially adjacent neighbors.

Maximization of $Q$ at a specific value of $\gamma$ is accomplished by varying $c$, the partition of network nodes into communities. In our context, this process results in the clustering of voxels into spatially contiguous communities whose members are more structurally similar to each other than the scaled average similarity given by $\gamma \langle W_{ij} \rangle$.
Each community is a ``crystal grain."
In practice, the algorithm that maximizes $Q$ is not guaranteed to find the global optimum \cite{Good2010}, so we perform modularity maximization 5 times and use the clustering that results in the maximum $Q$ value.

We provide intuition for how crystal grains subdivide white matter by showing a crystal grain partition $c$, obtained from modularity maximization at $\gamma=1.1$, in a slice through the example HARDI image described previously (Fig. \ref{fig:methods}D) 
Voxels are colored identically if they are members of the same crystal grain.
Grains reflect regions of white matter with similar structure; note the large grains that correspond to the corpus callosum, superior longitudinal fasciculus, and surrounding regions.
Note that here we use a single $\gamma$ resolution parameter for illustrative purposes. It remains of interest, however, to study the structure, morphology, and location of crystal grains over a range of resolutions. Accordingly, in a later section of this paper devoted to generating white matter parcellations according to crystal grain, we vary $\gamma$ and show that across a range of $\gamma$ values we obtain parcellations that are statistically similar to a white matter atlas currently used in the literature.

The white matter parcellation against which we choose to compare our results is the commonly referenced Johns Hopkins University (JHU) atlas \cite{Mori2005}, which was generated by hand segmentation of an average diffusion MRI tensor map over 81 adults.
We quantify the similarity between our parcellations over a range of $\gamma$ and the JHU atlas via two distinct measures of partition overlap, the adjusted RAND index \cite{Hubert1985} and the adjusted mutual information \cite{Vinh2010} between the partitions.
The RAND index between two partitions measures the fraction of pairs of elements in the system whose mutual classification is in agreement across both partitions -- that is, the fraction of element pairs determined in both partitions either to belong to the same cluster or to belong to different clusters.
The mutual information between two partitions gives the difference between the entropy of one partition and the conditional entropy of that same partition given the other; in other words, it measures the reduction in the information needed to encode one partition given that the other is known.
Adjusted versions of both measures correct for chance according to a permutation model, and normalize such that both measures are bounded above by 1, corresponding to exact agreement between partitions.
Calculations of both metrics were performed using the Python module \emph{scikit-learn} \cite{Pedregosa2011}.

\subsection{Other diffusion metrics}
We demonstrate the novelty of the information communicated by our crystallinity parameter $\tilde{\Delta}$ by comparing it to four common per-voxel structural metrics.
Here, we briefly describe these parameters: generalized fractional anisotropy (GFA), fractional aniostropy determined from diffusion tensor fitting (DTI-FA), the isotropic diffusion component (ISO), and mean diffusivity (MD).
GFA, the standard deviation of the ODF signal over its entire surface (normalized by the root-mean-square of the signal over its surface), is a unitless parameter that indicates the degree of anisotropy in the diffusion signal \cite{Tuch2004}.
DTI-FA is very similar in spirit to GFA; it is the standard deviation of the eigenvalues of the diffusion tensor, normalized by their root-mean-square \cite{Jones2013}.
ISO, the isotropic part of the spin ODF signal, is simply the minimum value of the diffusion signal in each voxel \cite{Yeh2010b}.
MD, the mean of the eigenvalues of the diffusion tensor in each voxel, characterizes the overall strength of the diffusion signal \cite{Jones2013}.

\subsection{Test-retest reliability and variability}
We confirm that our crystallinity metric $\tilde{\Delta}$ is well-behaved as a structural marker by calculating statistical proxies for reliability, reproducibility, and across-individual variability of $\tilde{\Delta}$ and other diffusion metrics.\\

\noindent \textbf{Reliability.} 
To measure reliability, we use two variants of the intra-class correlation coefficient ($ICC$). 
In general, the ICC reports variance of interest (in our case, variance of the metric across subjects) as a fraction of total variance.
When this fraction is high, the metric reliably differentiates individuals from each other, because variance across individuals is the majority contributor to the total variance.
We consider both per-voxel and per-image measures of $ICC$; the former is useful to indicate the spatial variance of reliability throughout the brain, while the latter condenses the reliability of the metric over an entire scan into one easily-interpretable scalar.

The per-voxel $ICC$ variant that we calculate, usually denoted by $ICC(3,1)$ \cite{Shrout1979}, uses a model with fixed session effects and random subject effects to describe the value of each metric in each voxel; in other words, total variance does not include effects due purely to across-session variability.
The quantity $ICC(3,1)$ can thus be thought of as a measure of consistency across scans, rather than absolute agreement across scans \cite{McGraw1996}.

We calculate $ICC(3,1)$ for each voxel as follows:

\begin{align}
    ICC(3,1) = \frac{MS_{BS} - MS_E}{MS_{BS}+(k-1)MS_E}.
\end{align}

\noindent Here $MS$ refers to the mean sums of squares, which we measure using a two-way analysis of variance (ANOVA). Specifically, $MS_{BS}$ is the between-subject mean sum of squares, and $MS_E$ is the residual mean sum of squares, after accounting for between-subject and between-measurement sums of squares. The quantity $k=8$ is the number of scans per subject.

We use the image intra-class correlation coefficient ($I2C2$) \cite{Shou2013} as a second, per-image measure of reliability.
The $I2C2$ is a generalization of the $ICC$:

\begin{align}
    I2C2 = 1 - \frac{Tr(K_U)}{Tr(K_W)}.
\end{align}

\noindent Here $K_W$ is the covariance matrix of the image vector $W$, where each entry of $W$ is the value of the metric in each voxel, concatenated over all subjects and all sessions.
The matrix $K_U$ is the covariance matrix of the measurement error vector $U$ over voxels, concatenated over all subjects and all sessions.
We calculate $I2C2$ for all metrics using the R package that accompanies Ref. \citenum{Shou2013}. \\

\noindent \textbf{Reproducibility.} To measure reproducibility, or agreement of each metric across multiple scans of the same subject, we calculate the within-subject coefficient of variation, $CV_{WS}$, for each voxel:

\begin{align}
    CV_{WS} = \frac{1}{\mu} \sqrt{\frac{1}{n} \sum_i \frac{1}{k-1} \sum_j \left( X_{ij} - \bar{X_i} \right)^2}.
\end{align}

\noindent Here, $n$ is the number of subjects, $k$ is the number of scans per subject, $X_{ij}$ is the value of the metric measured in subject $i$ and scan $j$, $\bar{X_i} \equiv \sum_j X_{ij}/k$ is the across-scan average of the metric for subject $i$, and $\mu$ is the grand mean of $X_{ij}$ over all subjects and scans.
In practice, we calculate $CV_{WS}$ using a two-way ANOVA: $CV_{WS} = \frac{1}{\mu} \sqrt{MS_{WS}}$, where $MS_{WS}$ is the within-subject mean sum of squares.
When $CV_{WS}$ is low, within-subject variation is low compared to its mean, and reproducibility is high.\\

\noindent \textbf{Individual Variability.} We finally measure across-individual variability by calculating the between-subject coefficient of variation, $CV_{BS}$, for each voxel:

\begin{align}
    CV_{BS} = \frac{1}{\mu}\sqrt{\frac{1}{n-1} \sum_i \left( \bar{X_i} - \mu \right)^2}
\end{align}
All variables are the same as those defined for $CV_{WS}$.
In practice, we calculate $CV_{BS}$ again via a two-way ANOVA, and use $CV_{BS} = \frac{1}{\mu} \sqrt{MS_{BS}/k}$, where $MS_{BS}$ is the between-subject mean sum of squares.

\subsection{Orientational order parameters}
To describe diffusion signal shape within each voxel, we use orientational order parameters that were first proposed by Steinhardt \emph{et al.} \cite{Steinhardt1983} to characterize the symmetries of particle clusters in supercooled liquids and glasses.
Consider that the diffusion signal can be represented as a probability density distribution on the unit sphere.
This distribution, in turn, can be expressed as a linear combination of the spherical harmonics, a set of basis functions on the unit sphere:

\begin{align}
    f(\theta, \phi) = \sum_{l=0}^{\infty} \sum_{m=-l}^l q_{lm} Y_{lm}(\theta, \phi). \notag 
\end{align}

\noindent Here, $Y_{lm}$ is the spherical harmonic associated with angular momentum number $l$ and magnetic quantum number $m$, and $q_{lm}$ is the projection of $f(\theta, \phi)$ onto $Y_{lm}$.
Using the Fourier coefficients $q_{lm}$, we construct the following orientational order parameter (sometimes also called a Steinhardt order parameter in the soft matter literature) to characterize the symmetry of the distribution associated with a specific angular momentum number $l$:

\begin{align}
    Q_l = \sqrt{\frac{4\pi}{2l+1} \sum_m \vert q_{lm} \vert^2}.
\end{align}
It can be shown that this parameter is rotationally invariant; that is, it does not change under any rigid rotation of $f(\theta, \phi)$.
A finite collection of $Q_l$ parameters for different values of $l$ is thus a rotationally-invariant, dimension-reduced fingerprint for the associated $f(\theta, \phi)$ distribution.
Rotational invariance is a useful property for any shape descriptor, because it means the descriptor depends only on the details of the shape geometry and not on its relative orientation with respect to an arbitrary reference frame.

\begin{figure*}
\centering
\includegraphics[width=0.8\textwidth]{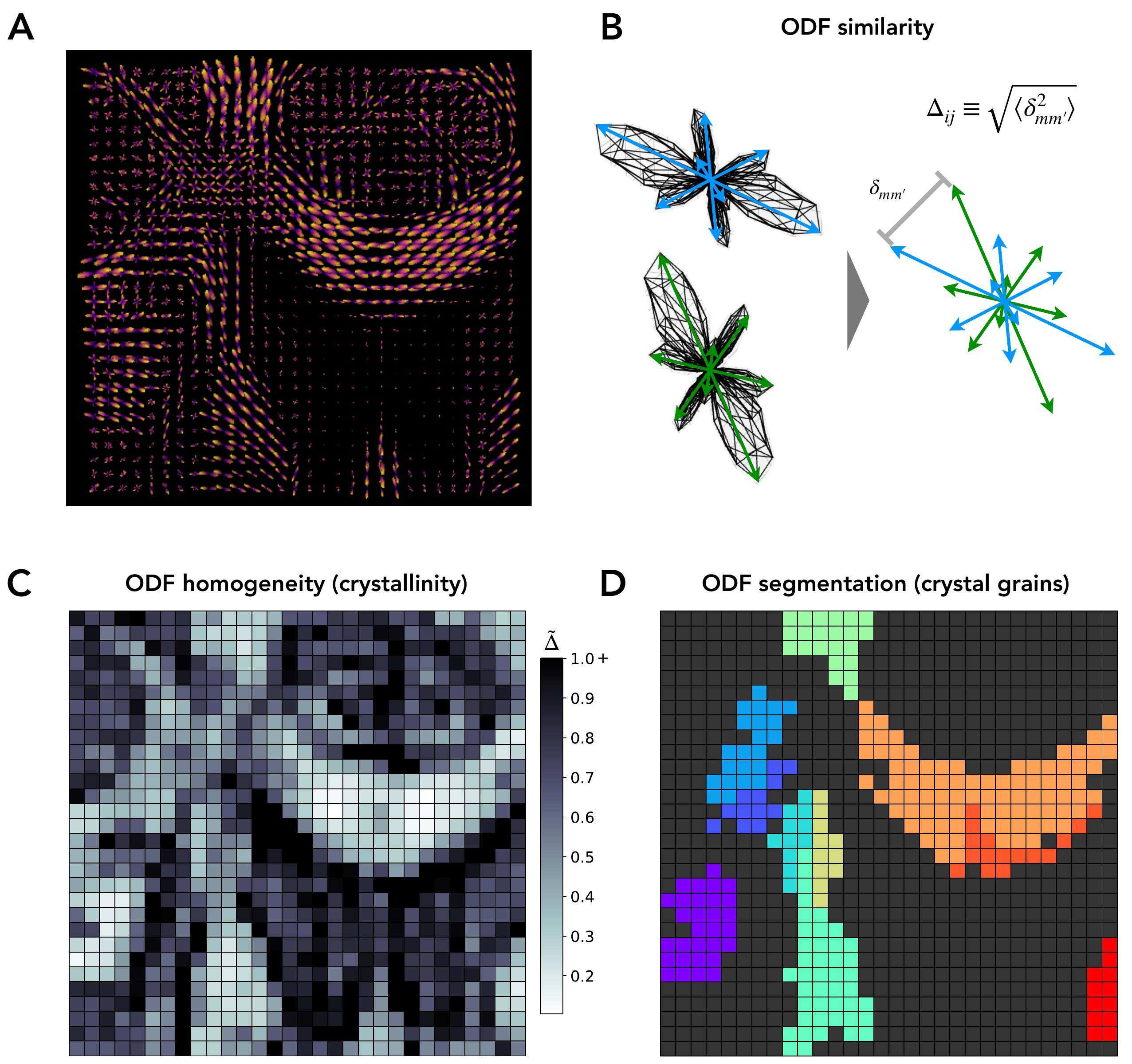}
\caption{\textbf{Crystallinity characterization methods used to analyze diffusion images of the human brain.}
(A) Slice of a sample diffusion image, showing the ODF signals reconstructed in each voxel.
(B) Schematic of our method for ODF similarity characterization via peak vector matching. 
Peak vectors are extracted from ODF distributions, and these vector sets are compared to find the best match between them, characterized by the lowest value of $\Delta_{ij}$.
(C) The identical diffusion image slice as that of panel (A), with voxels colored according to ODF homogeneity (crystallinity) according to the colorbar shown. 
Voxels with lower values of $\tilde{\Delta}$ are lighter (more crystalline), and voxels with higher values of $\tilde{\Delta}$ are darker (less crystalline).
(D) The identical diffusion image slice as that of panel (A), with voxels colored according to segmentation by crystal grain.
Voxels are colored identically if they are members of the same crystal grain.
Voxels colored black are members of crystal grains of size 10 voxels or less, or are not members of a crystal grain at all.
Otherwise, color has no additional meaning.
}
\label{fig:methods}
\end{figure*}

\section{Results}

\subsection{Crystallinity measures novel information} \label{section:hardi}
We first demonstrate that crystallinity varies across the brain in a meaningful and interpretable way, and provides new structural information that is not measured by other diffusion metrics commonly used in the neuroimaging community.
We calculate crystallinity in two separate datasets, and compare it to two other diffusion metrics, generalized fractional anisotropy (GFA) and mean diffusivity (MD), described in the \emph{Methods}.
We choose GFA and MD as our metrics of comparison because they measure very distinct aspects of the diffusion signal; thus, that crystallinity is distinct from each of them demonstrates its novelty.

\begin{figure*}
\centering
\includegraphics[width=0.9\textwidth]{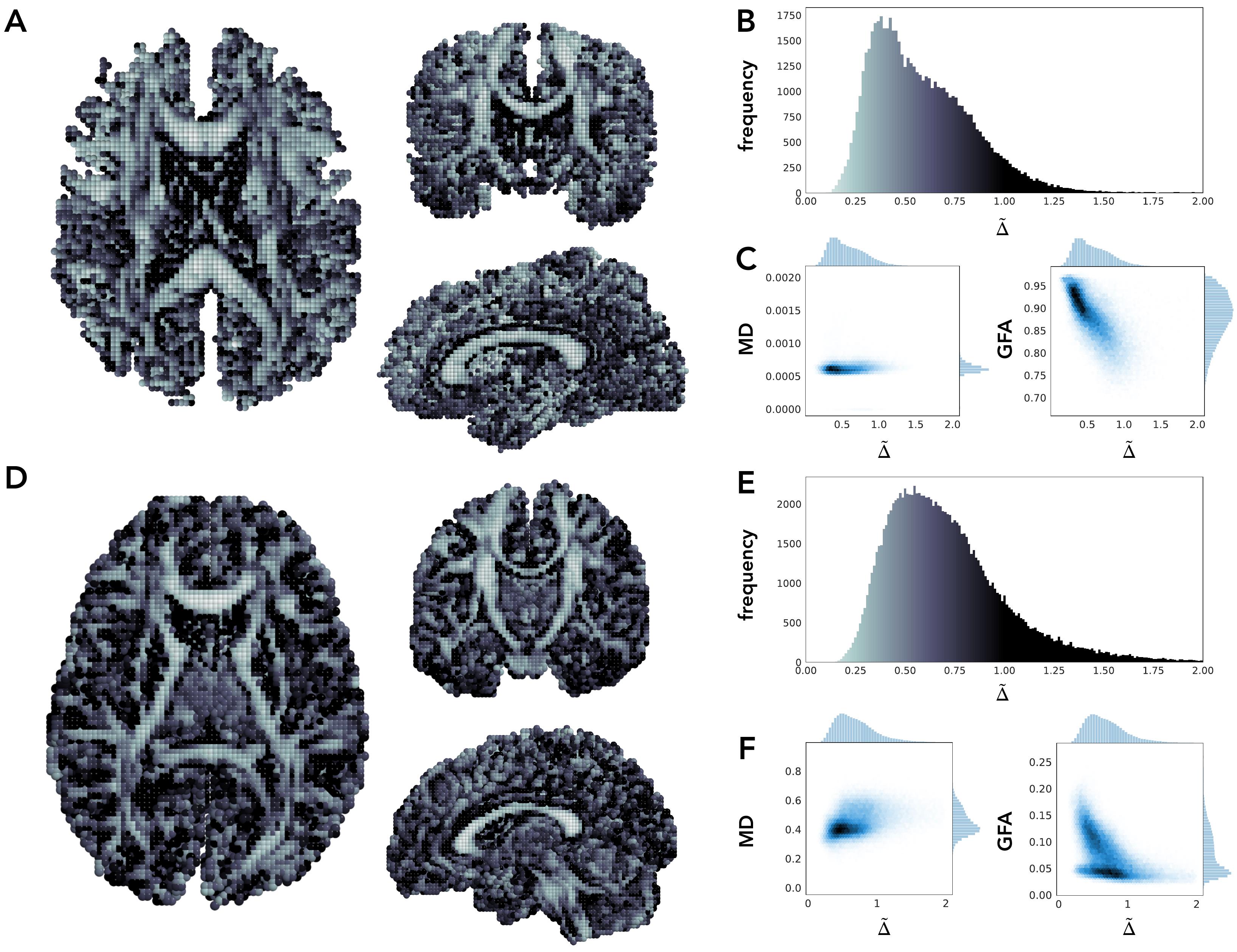}
\caption{\textbf{Crystallinity is sensitive to biological microstructure while providing novel information compared to common measures.}
(A) Three views of a sample diffusion image, where voxels are represented as spheres and colored according to $\tilde{\Delta}$.
(B) Histogram of the $\tilde{\Delta}$ values calculated for the voxels of this sample diffusion imaging scan.
Bars are colored to illustrate the coloring scheme used in panel (A). 
Voxels with lower values of $\tilde{\Delta}$ are lighter (more crystalline), and voxels with higher values of $\tilde{\Delta}$ are darker (less crystalline).
(C) Joint histograms of $\tilde{\Delta}$ against mean diffusivity (MD; left panel) as well as generalized fractional anisotropy (GFA; right panel), accumulated over the voxels of the sample scan.
(D) Three views of a session-averaged diffusion image for one subject in the CRASH study, with voxels colored according to $\tilde{\Delta}$ averaged across scans.
(E) Histogram of session-averaged $\tilde{\Delta}$ values calculated for the voxels of this subject.
Bars are colored to illustrate the coloring scheme used in panel (D).
(F) Joint histograms of $\tilde{\Delta}$ against mean diffusivity (MD; left panel) as well as generalized fractional anisotropy (GFA; right panel), accumulated over the voxels in one imaging session of the same subject.
}
\label{fig:hardi}
\end{figure*}

As an initial benchmark, we first compute the crystallinity of each voxel in the example HARDI image described in the \emph{Methods}.
Fig. \ref{fig:hardi}A shows three views of the HARDI image, with each voxel represented as a sphere and colored according to $\tilde{\Delta}$, and Fig. \ref{fig:hardi}B shows the corresponding histogram of $\tilde{\Delta}$.
Regions of high and low crystallinity clearly emerge, and the corpus callosum, corona radiata, internal capsule, and anterior commissure appear as regions of high crystallinity.
We directly compare crystallinity to common diffusion metrics by plotting joint distributions of $\tilde{\Delta}$ and MD across all voxels, and $\tilde{\Delta}$ and GFA across all voxels (Fig. \ref{fig:hardi}C).
The joint distributions show that crystallinity is not correlated with MD, and is somewhat correlated with GFA, since voxels of high GFA tend to be more crystalline, with lower values of $\tilde{\Delta}$.
These voxels tend to exist within white matter fascicles, where water diffuses in the same direction throughout neighboring voxels, and they thus have homogeneous structure.
In contrast, voxels with lower values of GFA have wider ranges of crystallinity, demonstrating that crystallinity provides information that is independent of diffusion anisotropy, especially in areas distinct from streamlined white matter.

To demonstrate that crystallinity provides novel information in real datasets, we also analyze diffusion images collected as part of the CRASH study described in the \emph{Methods}.
We take advantage of the multiple scans provided for each subject in this dataset to obtain smoother across-session averages of voxel-wise crystallinity, $\langle \tilde{\Delta} \rangle_{session}$, which we show in Figs. \ref{fig:hardi}D,E for one CRASH subject.
This crystallinity brain map is qualitatively consistent with the one calculated for the example HARDI image (Fig. \ref{fig:hardi}A): crystalline regions emerge throughout the brain, including the corpus callosum, corona radiata, internal capsule, and anterior commissure.
``Grain boundaries," or voxels of lower crystallinity on the boundary of regions of higher crystallinity, can also be seen.

We again compare crystallinity to the structural metrics MD and GFA via joint distributions across voxels. Fig. \ref{fig:hardi}F shows distributions for one scan of the subject represented in panels D and E.
We find similar relationships between the metrics as those found for the HARDI example (Fig. \ref{fig:hardi}C).
Namely, $\tilde{\Delta}$ is uncorrelated with MD and is lower for voxels of higher GFA.
The joint distribution of $\tilde{\Delta}$ and GFA for this CRASH scan shows, even more clearly than the distribution for the HARDI example, that crystallinity and GFA are increasingly independent metrics as GFA decreases.
Voxels of lower anisotropy have larger ranges of $\tilde{\Delta}$, and indeed an entire set of voxels of low GFA $\sim 0.05$ has the widest range of $\tilde{\Delta}$ values, between $\tilde{\Delta}=0$ and $\tilde{\Delta}=2$.
These voxels are primarily in non-white matter regions of the brain 
and thus have more isotropic diffusion signals.

\subsection{Test-retest reliability and variability across individuals} \label{section:reliability}
To confirm that our crystallinity metric $\tilde{\Delta}$ is a well-behaved structural marker whose variation is meaningful, we assess the reliability, reproducibility, and across-individual variability of $\tilde{\Delta}$ and the four common diffusion metrics described in the \emph{Methods}, and compare these quantities.
We find that crystallinity is reliable, fairly reproducible, and has higher across-individual variability than the other common diffusion metrics.
Our results together indicate that crystallinity may be useful to distinguish individual differences in white matter along the dimensions of sex, genetics, and physiology.

We measure all diffusion metrics in 8 scans each of 25 subjects in the CRASH study, and calculate statistical quantities for each metric that are proxies for reliability, reproducibility, and across-individual variability, in the spirit of prior work \cite{LuqueLaguna2020,bassett2011conserved,pfefferbaum2003replicability}.
Details of each statistical quantity can be found in the \emph{Methods}.
We then warp the scalar fields for each metric and each scan into a master template space via the nearest-neighbor interpolation method in ANTs \citep{ants}, to facilitate voxel-wise comparison across subjects.\\

\begin{figure*}
\centering
\includegraphics[width=0.8\textwidth]{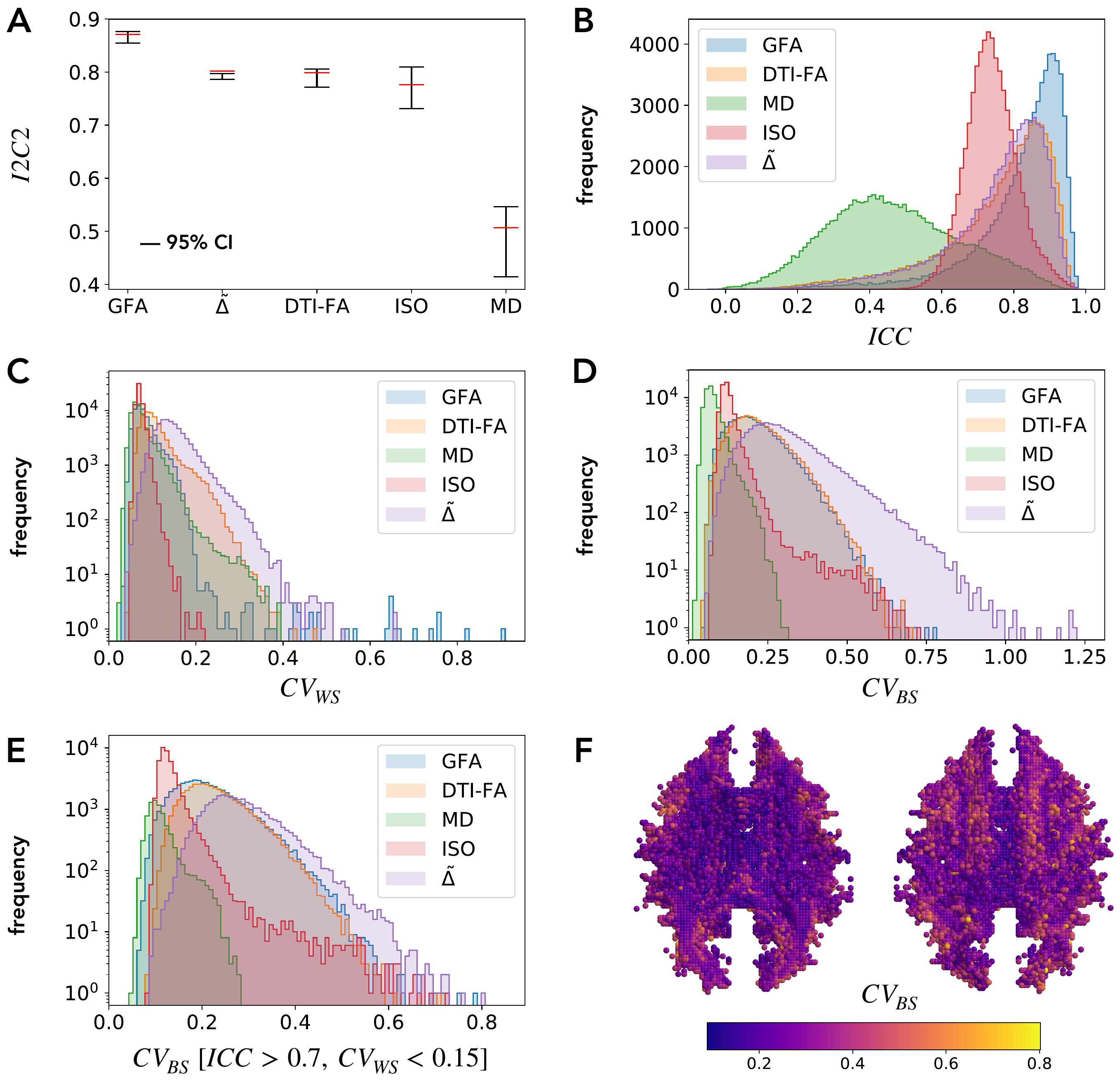}
\caption{\textbf{Crystallinity is reliable and reproducible across scans and has high inter-subject variability.}
(A) Image intra-class correlation coefficient ($I2C2$) calculated over the CRASH dataset for four common structural metrics as well as our metric ($\tilde{\Delta}$).
Error bars denote 95\% confidence intervals, determined via bootstrapping in which we resampled subjects 50 times with replacement and recalculated $I2C2$.
(B) Histograms of intra-class correlation coefficient $ICC(3,1)$ (denoted $ICC$ in the panel) over all voxels of the identical dataset.
(C) Histograms of the within-subject coefficient of variation, $CV_{WS}$, over all voxels of the identical dataset.
(D) Histograms of the between-subject coefficient of variation, $CV_{BS}$, over all voxels of the identical dataset.
(E) Histograms of $CV_{BS}$ for only those voxels with reliable and reasonably reproducible values of each metric, determined by $ICC(3,1) > 0.7$ and $CV_{WS} < 0.15$.
(F) Two views of a map of $CV_{BS}$ for $\tilde{\Delta}$ for the voxels shown in panel (E).
Voxels are colored by $CV_{BS}$ according to the color bar below the images.
The view on the left shows an axial slice of the image, and the view on the right is the flipside of the view on the left, showing the exterior of the image.
}
\label{fig:reliability}
\end{figure*}

\noindent \textbf{Reliability.} We find that $\tilde{\Delta}$ is very reliable across subjects according to both $ICC(3,1)$ and $I2C2$ (Figs. \ref{fig:reliability}A,B).
The mean value of $I2C2$ for $\tilde{\Delta}$ across all subjects is $\sim 0.8$, only lower than the I2C2 for GFA (Fig. \ref{fig:reliability}A).
Similarly, the distribution of ICC across all voxels in the common group mask is peaked above 0.8 for $\tilde{\Delta}$, in approximately the same location as the ICC distribution for DTI-FA.
For reference, we note that values of $ICC > 0.7$ imply high reliability in other imaging studies \cite{Marenco2006, Somandepalli2015, LuqueLaguna2020}.
Only the ICC distribution for GFA is peaked at a significantly higher ICC value (Fig. \ref{fig:reliability}B).\\

\noindent \textbf{Reproducibility.} We find that our metric $\tilde{\Delta}$ is fairly reproducible, with $CV_{WS}$ peaked around 0.13 (Fig. \ref{fig:reliability}C).
In other imaging studies, values of $CV_{WS} < 0.1$ imply high reproducibility \cite{Marenco2006,LuqueLaguna2020}
We note, however, that the location of this peak is highest for $\tilde{\Delta}$ among all the metrics analyzed, and that more voxels have $CV_{WS} > 0.13$ for $\tilde{\Delta}$ than for the other metrics.
This result indicates that $\tilde{\Delta}$, although a reliable differentiator of individuals, is slightly less reproducible across scans than other metrics such as GFA and ISO.\\

\noindent \textbf{Individual Variability.} The shape of the $CV_{BS}$ distribution for $\tilde{\Delta}$ has a significantly longer tail than the distributions for the other metrics, implying that the crystallinity of a large subset of voxels is more variable across subjects than the other metrics analyzed (Fig. \ref{fig:reliability}D).
Even within the more restrictive subset of voxels that are reasonably reproducible, with $CV_{WS} < 0.15$, and reliable, with $ICC(3,1) > 0.7$, the individual variability of $\tilde{\Delta}$ is high in comparison with the other diffusion metrics (Fig. \ref{fig:reliability}E).
The distribution of $CV_{BS}$ for these voxels still has a longer tail and a peak at larger $CV_{BS}$ for $\tilde{\Delta}$ than the other metrics, although these features are less pronounced than in Fig. \ref{fig:reliability}D.
A map of $CV_{BS}$ for $\tilde{\Delta}$ over these voxels (Fig. \ref{fig:reliability}F) shows that voxels with reliable, fairly reproducible, and across-subject variable values of $\tilde{\Delta}$ tend to be located on the gray matter and lateral surfaces, rather than the white matter and medial surfaces.

\subsection{Crystalline regions of the brain} \label{section:xtalWM}
We now turn to a more thorough investigation of crystallinity throughout the brain, again via analysis of the scans collected in the CRASH study. 
We examine the spatial distribution of crystallinity in two ways. 
First, we generate a voxel map of crystallinity, averaged over all subjects and all scans.
We also segment each scan into neuroanatomical regions using a common white matter parcellation, and measure crystallinity throughout each region.
In general, we find that crystalline voxels are located towards the center of the brain, in the carona radiata, internal capsule, corpus callosum, and uncinate fasciculus.
Our findings are consistent with the results reported in Section \ref{section:hardi}.

\begin{figure*}
\centering
\includegraphics[width=0.9\textwidth]{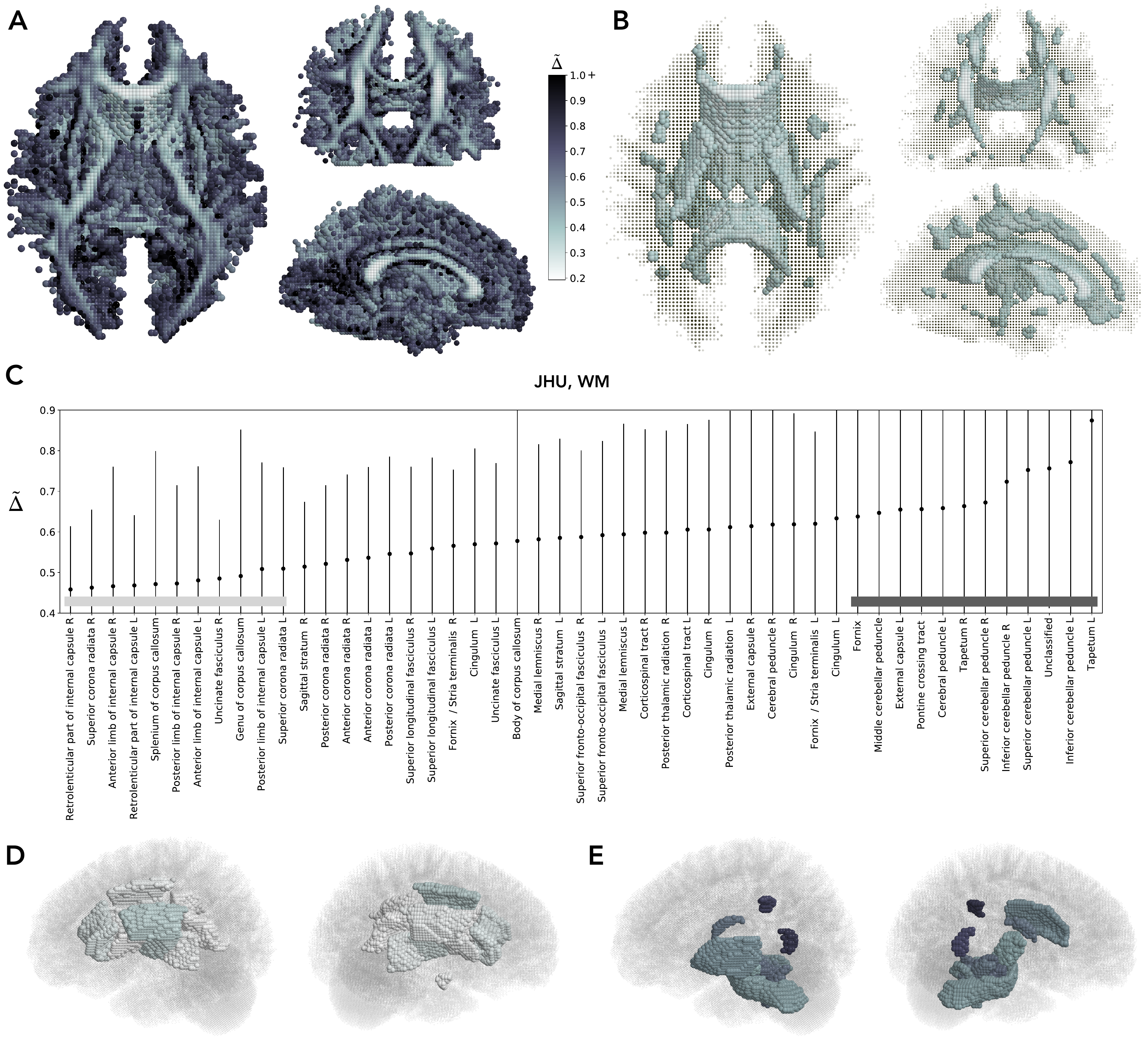}
\caption{\textbf{Identification of crystalline regions of white matter.}
(A) Three views of the crystallinity proxy $\tilde{\Delta}$ averaged over all subjects and all scans of the CRASH dataset, shown in a master template space as described in the main text.
Each voxel is colored by average $\tilde{\Delta}$ according to the color bar to the right of the images. 
Voxels with lower values of $\tilde{\Delta}$ are lighter (more crystalline), and voxels with higher values of $\tilde{\Delta}$ are darker (less crystalline).
(B) Thresholded views identical to those shown in panel (A), where only the (most crystalline) voxels with the lowest 10\% of $\tilde{\Delta}$ values are shown.
Voxels are also colored according to the color bar of panel (A).
(C) Crystallinity proxy $\tilde{\Delta}$ of voxels in white matter regions according to the JHU white matter atlas (described in the main text), pooled over all subjects and all sessions.
White matter regions are denoted with their formal anatomical names, and circles show the mean value of $\tilde{\Delta}$ in each region, taken across all subjects and all sessions.
Error bars show the standard deviation of $\tilde{\Delta}$ within each region.
The plot is shown zoomed in for clarity, and regions are organized by increasing mean $\tilde{\Delta}$.
In some cases, error bars stretch outside of boundaries of this zoomed-in view.
(D,E) The most (D) and least (E) crystalline white matter regions in a sample scan, according to the pooled results over subjects and sessions shown in panel (C).
The most crystalline regions are marked with a light gray bar in panel (C), and the least crystalline regions are marked with a dark gray bar.
Each region is colored according to its average value of $\tilde{\Delta}$.
}
\label{fig:xtalWM}
\end{figure*}

The subject-averaged spatial distribution of crystallinity throughout the brain, in the master template space discussed in Section \ref{section:reliability}, is shown in Fig. \ref{fig:xtalWM}A.
Each voxel is colored according to its value of $\langle \tilde{\Delta} \rangle_{scan,subj}$, the average value of $\tilde{\Delta}$ over all scans for all subjects.
Fig. \ref{fig:xtalWM}B contains a thresholded version of the map in panel (A), showing only voxels with the lowest 10\% of $\tilde{\Delta}$ values.
The averaged crystallinity varies smoothly throughout the brain, and more crystalline regions of the brain are localized in white matter than in grey matter.

To gain more insight into the crystallinity of specific regions of the brain, we utilize a common white matter parcellation, the JHU atlas described in the \emph{Methods}, to group voxels into neuroanatomical regions. 
We then calculate distributions of $\tilde{\Delta}$ in each white matter region, pooled over all subjects and all scans.
In contrast to the analysis presented previously, we warp the JHU white matter atlas into each subject's template space in this analysis, to utilize all the data collected during each scan.
To perform the warping, we use the genericlabel interpolation method in ANTs \cite{ants}.
Fig. \ref{fig:xtalWM}C shows distributions of $\tilde{\Delta}$ in each white matter region according to the JHU atlas.

We highlight regions of especially low $\tilde{\Delta}$ (high crystallinity) and high $\tilde{\Delta}$ (low crystallinity) with colored bars. We indicate the locations of these regions in a sample scan in Figs. \ref{fig:xtalWM}D,E.
Highly crystalline regions include the internal capsule, splenium and genu of the corpus callosum, carona radiata, and uncinate fasciculus. Regions of low crystallinity are located in the fornix, tapetum, and cerebellar peduncle.

\subsection{Parcellation according to crystal grain} \label{section:clustering}
Finally, we explore the parcellation of white matter into ``crystal grains," or spatially contiguous communities of voxels that are structurally more similar to each other than expected on average (see the \emph{Methods}).
We find that, despite its anatomy-blind automation, our crystal grain parcellation still results in white matter regions of anatomical interest, and may be a useful way of segmenting white matter in future studies.

\begin{figure*}
\centering
\includegraphics[width=0.9\textwidth]{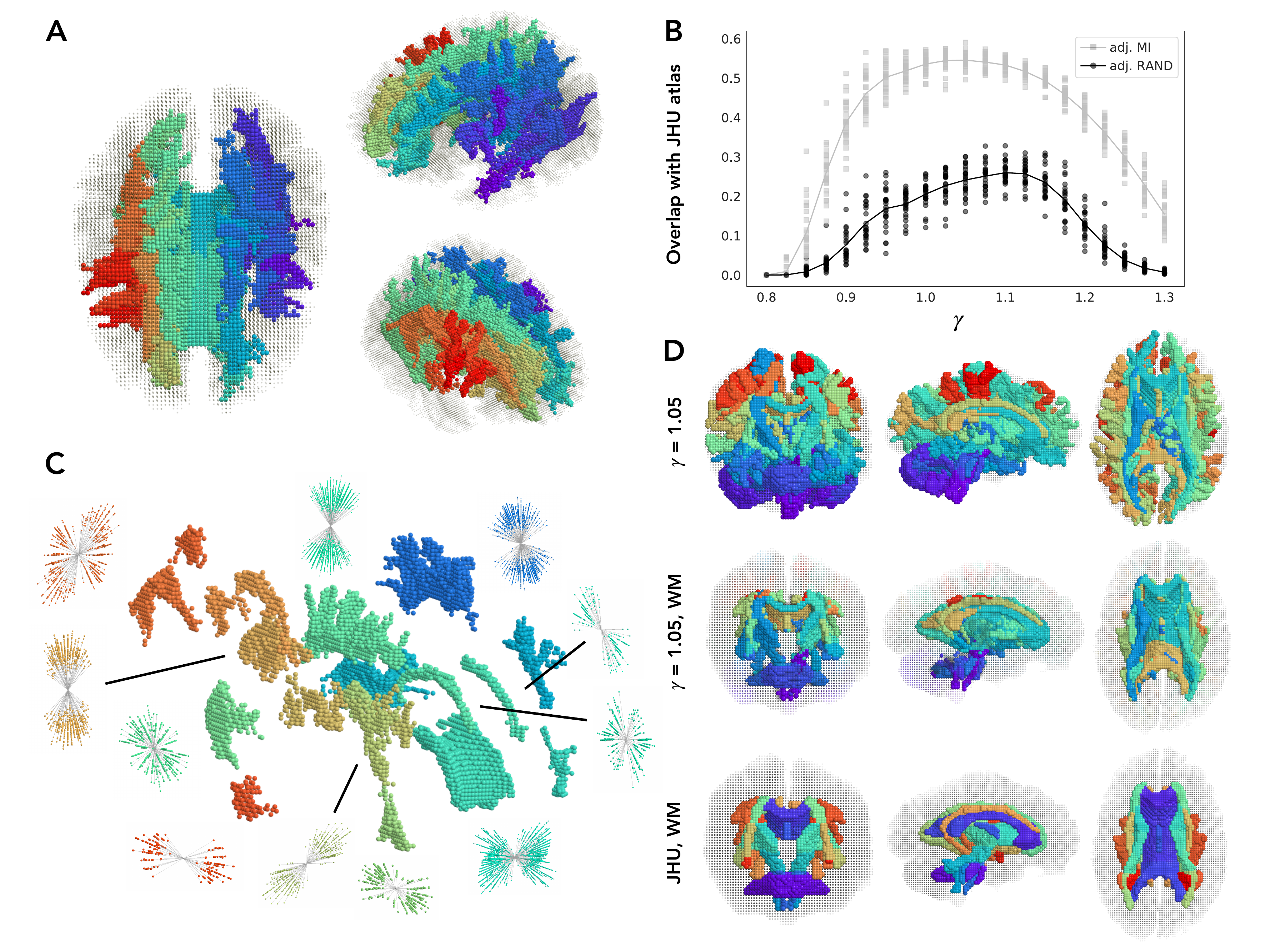}
\caption{\textbf{Parcellation of white matter into clusters of voxels with similar structure, or ``crystal grains."}
(A) Three views of a sample diffusion image, colored according to crystal grain.
Voxels are colored identically if they are members of the same crystal grain.
Voxels colored black (and smaller in size for ease of visibility) are members of crystal grains of size 100 voxels or less.
(B) Measures of partition similarity between our segmentation according to crystal grain and the common JHU white matter atlas, computed over scans in the CRASH dataset.
Adjusted mutual information (adj. MI) is shown in gray and adjusted RAND score (adj. RAND) is shown in black as a function of community resolution parameter $\gamma$.
Each marker is a measure of partition similarity for one subject, averaged over its sessions, and lines denote averages across subjects.
Similarity is significant over a range of $\gamma$, and peaks especially around $\gamma \sim 1.0-1.1$.
(C) Select crystal grains shown in panel (A), with associated composite diffusion signals.
Each composite signal is colored identically to its associated crystal grain, and connecting lines are drawn when the association is not clear.
Composite diffusion signals are represented via ball-and-stick models for all peak vectors, and peak vectors for all voxels in each crystal grain are shown simultaneously.
(D) Three views of our crystal grain parcellation of the subject with highest adjusted mutual information with the JHU atlas, at $\gamma = 1.05$.
Voxels are colored identically if they are members of the same crystal grain of size $>$ 100 voxels.
The top row of images shows the full crystal grain segmentation, while the middle row shows only those voxels in white matter regions according to the JHU atlas.
The JHU atlas itself is shown in the bottom row of images for comparison; each white matter region in the atlas is colored an identical, random color.
Similar anatomical structures can be seen in each white matter parcellation.
}
\label{fig:cluster}
\end{figure*}

We first show an example parcellation by crystal grain in Figs. \ref{fig:cluster}A-C to demonstrate that crystal grains are anatomically meaningful by eye and contain useful structural information.
We consider the same diffusion image as that analyzed in Section \ref{section:hardi}, and color crystal grains of size 100 voxels or greater, detected using community resolution parameter $\gamma=1.1$. (Fig. \ref{fig:cluster}A).
Crystal grains are roughly hemispherically symmetric and follow the anatomical branching of large white matter tracts.
A closer look at grains deep within the brain near the corpus callosum (Fig. \ref{fig:cluster}C) shows the hemispherical symmetry further, as well as the composite structural environments of individual grains.

We next compute crystal grains in all scans of the CRASH dataset, and compare pooled results from our parcellation scheme to the commonly used JHU white matter atlas introduced in the \emph{Methods}.
We perform crystal grain segmentation over a weighted network for each subject, where weights are defined as in the \emph{Methods} and averaged over all sessions of the subject.
We vary the community resolution parameter $\gamma$ for each subject, and find significant overlap between our crystal grain parcellation and the JHU atlas over a range of $\gamma$ for all subjects (Fig. \ref{fig:cluster}B).
Partition overlap is measured via two distinct metrics, the adjusted RAND index (adj. RAND) and adjusted mutual information (adj. MI), both described in the \emph{Methods}, to show that trends in overlap with $\gamma$ are general and do not depend on the specific overlap measure.
Both overlap measures peak between $\gamma=1.0$ and $\gamma=1.1$.
Visual inspection of our crystal grain parcellation of the subject with highest adjusted mutual information with the JHU atlas (at $\gamma = 1.05$) confirms that the segmentation is in fact similar to the JHU atlas (Fig. \ref{fig:cluster}D).
We find mesoscale crystal grains that resemble white matter regions in the JHU atlas, yet are completely determined by automatic detection of homogeneous brain tissue architecture. Our results underscore the potential for crystallinity to provide a data-driven microstructural feature for white matter parcellation.

\subsection{Diffusion signal shape characterization}
While crystallinity provides important information regarding the homogeneity of diffusion signals within a local region of voxels, characterization of the diffusion signal shape itself within each voxel is also important for locating white matter fiber crossings and fascicles.
For this purpose, tools from soft matter may also be useful; we briefly explore one such structural characterization tool here. 
We use sets of orientational order parameters-- usually denoted by $Q_l$ in soft matter literature and used to describe symmetries of particle clusters \cite{Steinhardt1983}-- to describe the shape of each diffusion signal in each voxel, and show that these $Q_l$ shape descriptors provide an easily calculable, succinct means of determining whether the diffusion signal shape is especially unidirectional (representing a coherent white matter fiber bundle), bidirectional (representing a fiber crossing), or noisy. 
Each diffusion signal shape can be described with a shape descriptor consisting of three orientational order parameters ($Q_2$, $Q_4$, $Q_6$); this acts as a fingerprint (of only three dimensions) for the shape of the signal that is quite distinct from other diffusion metrics like MD and GFA.
Calculation of the $Q_l$ parameters is described in the \emph{Methods}. 

\begin{figure*}
\centering
\includegraphics[width=0.8\textwidth]{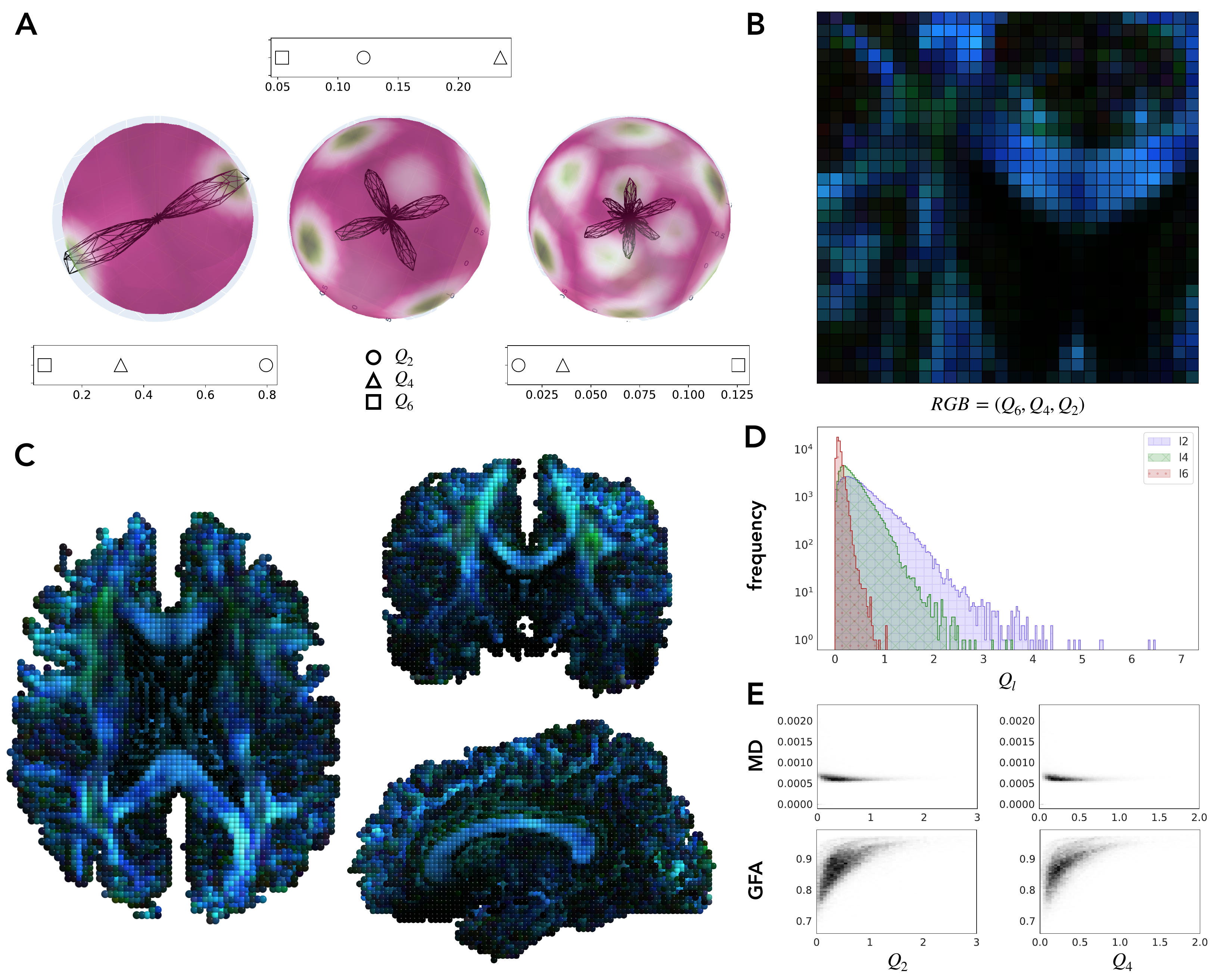}
\caption{\textbf{Orientational order parameters characterize diffusion signal shape.}
(A) Three fiber ODFs with different shapes are shown in black at the center of their corresponding probability density distributions on the unit sphere. 
Higher probability density is colored green, and lower probability density is colored pink. 
Above or below each ODF, corresponding values of orientational order parameters $Q_2$ (circles), $Q_4$ (triangles), and $Q_6$ (squares) are shown.
(B) A slice of an example HARDI image, with voxels colored according to orientational order parameters $Q_2$, $Q_4$, and $Q_6$. 
For each voxel, ($Q_6$, $Q_4$, $Q_2$) is converted to an (R,G,B) color according to $\min(Q_l, 1)$.
(C) Three views of a full sample image, with voxels colored according to orientational parameters ($Q_6$, $Q_4$, $Q_2$) = (R,G,B). Each $Q_l$ parameter is converted to the appropriate color channel according to $\min(Q_l, 1.25)$. 
Voxels that appear cyan are those for whom $Q_2 > 1.25$ and $Q_4 > 1.25$, resulting in equal mixing of the blue and green channels.
(D) Histograms of $Q_l$ for all voxels of this sample scan for $l=2$ (blue vertical cross-hatch), $l=4$ (green diagonal cross-hatch), and $l=6$ (red dots).
(E) Joint histograms of $Q_2$ (left) and $Q_4$ (right) against MD (top) and GFA (bottom) accumulated over all voxels of this sample scan.
}
\label{fig:Ql}
\end{figure*}

To show that the set ($Q_2$, $Q_4$, $Q_6$) differentiates between various diffusion signal shapes, we show three examples of fiber ODFs (unidirectional, crossing, and noisy), their corresponding probability density distributions on the unit sphere, and values of orientational order parameters $Q_2$, $Q_4$, and $Q_6$ for each of them (Fig. \ref{fig:Ql}A).
The especially unidirectional ODF's value of $Q_2$ is significantly higher than the other order parameters; the crossing ODF's value of $Q_4$ is significantly higher than the other order parameters; and the noisier ODF's value of $Q_6$ is highest, although absolute values for all parameters are different across ODFs.
Viewing the parameters ($Q_2$, $Q_4$, $Q_6$) together thus communicates significant information regarding diffusion streamlines, crossings, and general noise.

We demonstrate the ability of the parameters ($Q_2$, $Q_4$, $Q_6$) to distinguish between streamlines, crossings, and noise in a real dataset by visualizing them in the example Stanford HARDI image discussed previously (Fig. \ref{fig:Ql}B,C).
We visualize ($Q_2$, $Q_4$, $Q_6$) simultaneously as blue, green, and red channels respectively when coloring each voxel.
Voxels with ODFs that are mostly uniaxial are mostly blue in color, as their values of $Q_2$ are highest.
Voxels with ODFs that indicate fiber crossings and therefore are biaxial are more green, as they contain higher values of $Q_4$.
Darker voxels contain noisier ODFs, as all values of $Q_l$ are lower in magnitude.
Although $Q_6$ is generally higher in these noisier voxels, they are not significantly red in color, because all $Q_l$ values in these voxels are low.
In both the diffusion imaging slice in Fig. \ref{fig:Ql}B and the three views of the whole scan shown in Fig. \ref{fig:Ql}C (each identical to the views shown in Fig. \ref{fig:methods}A and Fig. \ref{fig:hardi}A respectively), the corpus callosum and corona radiata emerge as a regions of a more blue or cyan color, indicating highly structured white matter fascicles, and more green regions of crossing fibers can also be seen at the juncture between large white matter tracts, particularly in the transverse section.
Cumulative distributions of each $Q_l$ parameter over the entire scan show the wider range of $Q_2$ with respect to the other parameters, and the wider range of $Q_4$ with respect to $Q_6$ (Fig. \ref{fig:Ql}D).

We also show that the orientational order parameters $Q_l$ contain novel information concerning each diffusion ODF, by directly comparing values of $Q_2$ and $Q_4$ of each voxel against corresponding values of mean diffusivity (MD) and generalized fractional anisotropy (GFA) (Fig. \ref{fig:Ql}E). 
We find that the orientational $Q_l$ parameters are not particularly correlated with MD, and are somewhat correlated with GFA, with the general trend that voxels of lower GFA tend to have lower values of $Q_2$ or $Q_4$.
As GFA increases, however, voxels have wider ranges of both $Q_2$ and $Q_4$, illustrating the decoupling of orientational diffusion shape information from anisotropy.

\section{Discussion}
We have demonstrated that structural characterization techniques popular in soft matter are also useful in the context of neuroimaging.
Crystallinity, or diffusion signal homogeneity, is a promising marker of white matter microstructure. 
It provides white matter information that is independent of other traditional diffusion metrics; it is reliable, reproducible, and variable across individuals; and it varies meaningfully throughout the brain, with large white matter bundles possessing more structural homogeneity than smaller white matter bundles or grey matter.
Segmentation of white matter into crystal grains, or especially structurally similar regions, constitutes a parcellation informed by underlying brain tissue architecture that overlaps significantly with a commonly used white matter atlas.
Finally, orientational order parameters from soft matter provide low-dimensional descriptors of diffusion signal shape, and readily pinpoint regions of white matter streamlines and fiber crossings.
The following sections provide context for our work within the separate disciplines of neuroimaging and soft matter, consider the limitations of our methods, and outline possible future directions and extensions.

\subsection{Prior methods to assess white matter microstructure}
A major obstacle for dMRI is its high dimensionality. Most popular neuroimaging methods result in a scalar value for each voxel that can be used in voxel-wise group analyses with standard statistical tools. Measures like (G)FA and MD simplify an ODF into a scalar value, losing information about its directionality and dispersion. Additionally, these values are difficult to interpret in biological terms \citep{jones2013a}. Alternatively, fixel-based analyses \citep{fixels} compare scalar values on each ODF lobe, providing greater specificity than voxel-based approaches. Fixels, however, do not consider the orientation of fixels relative to one another or across voxels. The approach proposed here incorporates magnitude and direction \emph{across voxels} to characterize structure.

\subsection{Common tools for structural characterization in soft matter}
Structural characterization is a vibrant subfield of soft matter physics, due to soft materials' susceptibility to thermal fluctuation and correspondingly messy and rearranging local configurations.
Identification of local structure is used to track nucleation and growth during material phase transitions from the fluid to the crystal phase, to identify crystalline defects or grain boundaries within solids, to identify meaningful local structures that may influence dynamics in amorphous systems, and even to identify crystalline symmetries in an automated fashion.
The methods to assess white matter microstructure that we have presented here, namely local crystallinity and orientational order parameterized by $Q_l$, are only two of the many techniques developed over the past several decades for structural characterization within soft matter \cite{Keys2011, Stukowski2012}.
These methods use local signatures ranging from Voronoi cells \cite{Finney1970a,Tanemura1977}, to bond angles \cite{Ackland2006} or the topology of local bonding environments \cite{Honeycutt1987,Malins2013b,Lazar2015}, to the Fourier coefficients of local bonding environments \cite{Steinhardt1983,Auer2004,Phillips2013,Spellings2018} to identify local structure in particulate systems.
Recent advancements have also been made in using machine learning to identify structures robustly and automatically \cite{Phillips2013,Reinhart2017,Spellings2018}.
Our method for detecting local crystallinity is most similar in spirit to polyhedral template matching \cite{Larsen2016}, which also relies on root-mean-squared deviations between local environments in real space to determine structure.
However, this method and many of those just described seek to identify specific structures by comparing them to candidate templates; our simpler method, by contrast, is agnostic with respect to structure.
We were solely concerned in this paper with structural \emph{homogeneity}, rather than the type of structure itself.

\subsection{Prior efforts to develop white matter parcellations}
Parcellation, or the division of the brain into meaningful regions, is a critical step in understanding and summarizing grey matter. 
Due to its importance, methods for the parcellation of grey matter in individual brains are widely used and available in common software packages such as Freesurfer.
Subcortical white matter atlases are also very useful for distinguishing neuroanatomical structures and providing regions of interest (ROIs) for cross-subject comparisons \cite{Smith2013}. 
However, white matter parcellation methods have proven difficult to automate, and prior approaches have relied on manual tract segmentation by neuroanatomists, either of physical post-mortem brains \cite{Eickhoff2005} or averages over diffusion images \cite{Mori2005}.
By contrast, our proposed classification of voxels into structurally similar regions in individual brains has the dual advantages of automation and a physically intuitive criterion for parcellation: ``crystal grains" in the brain are agnostic to neuroanatomy and depend only on structural similarity.

\subsection{Methodological considerations and limitations}
Our method of crystallinity characterization is limited in several ways, and may be improved in the future.
First, we only use the peaks of each ODF signal to characterize the local white matter environment.
Although this focus has the advantage of reducing the dimensionality needed to describe each ODF signal, it obviously results in a large loss of signal information.
Future efforts to determine structural homogeneity could rely on other shape matching methods that incorporate more information about each signal shape.
Additionally, we compared structural environments using a greedy method to minimize their root-mean-squared deviation.
This method, while fast and straight-forward, is not guaranteed to find the globally minimum deviation between the environments.
Finding the global minimum amounts to solving the well-known assignment problem \cite{Aardal2005}, and in the future we could replace our greedy algorithm with any of the algorithmic solutions to the assignment problem to compare structural environments and determine crystallinity.
Finally, we note that we have characterized the crystallinity of each voxel via a metric that is normalized with respect to the signal strength of that voxel. 
We included this normalization to treat noisy signals of low strength on equal footing with clear signals of high strength, but a better normalization scheme in the future could be developed to actively weight noisy signals such that they are explicitly classified as disordered.

\subsection{Future directions}
A study in the immediate future will use the orientational order parameters $Q_l$ introduced in this work to characterize white matter fiber crossings and streamlines in real datasets such as the CRASH dataset, to ascertain the robustness and reliability of this characterization method, and to investigate differences in these parameters across individuals.
We will also explore using the rotational invariance of these parameters for the purposes of brain registration and image correction.

We also look forward to using local crystallinity to investigate individual differences in brain structure that may occur as a result of sex, genetics, or physiology.
Future studies will explore how structural homogeneity changes in white matter during development, and if it varies with gene expression in distributed regions throughout the brain.
For each dimension of variance, we will target questions such as: 
Do associated statistics involving overall crystallinity change, or are shifts localized to certain regions? 
Which regions become more or less crystalline? 
Does crystallinity correlate with any other biomarkers that we can glean from pertinent datasets?
Additional tools from the soft matter community may also be leveraged to characterize local white matter structure in more detail, including automated structure classification via machine learning.
These investigations will contribute to the growing body of knowledge regarding how subcortical structure influences brain function.

\section{Conclusion}
Materials science provides advanced approaches for describing the characteristics of soft matter, but these tools have not yet been used to understand the complex structure of the human brain.
This study provides an initial step in that interdisciplinary direction.
We have analyzed white matter microstructure in the human brain by calculating crystallinity, or diffusion signal homogeneity, and orientational order parameters that describe the shape of each diffusion signal.
We found that crystallinity provides information independent of that of other common diffusion metrics, has high test-retest reliability, and varies meaningfully across the brain along interpretable lines of anatomical difference.
Parcellations of white matter into crystal grains, or regions with high structural similarity, are automated, informed by brain architecture, and have significant overlap with a common white matter atlas.
Additionally, sets of orientational order parameters provide fingerprints of reduced dimensionality for each diffusion signal that are capable of locating regions of white matter fiber crossings and streamlines.
We hope that our paper inspires future communication between the fields of soft matter physics and neuroimaging, and that the work presented here is only the beginning of many fruitful collaborations.

\section{Citation diversity statement}
Recent work in several fields of science has identified a bias in citation practices such that papers from women and other minorities are under-cited relative to other papers in the field \cite{Dworkin2020, maliniak2013gender, caplar2017quantitative, chakravartty2018communicationsowhite, YannikThiemKrisF.SealeyAmyE.FerrerAdrielM.Trott2018, dion2018gendered, bertolero2020racial}. 
Here we sought to proactively consider choosing references that reflect the diversity of our field in thought, form of contribution, gender, and other factors.
We obtained the predicted gender of the first and last author of each reference by using databases that store the probability of a first name being carried by a woman or a man \cite{Dworkin2020,cleanbib}. 
By this measure (and excluding self-citations to the first and last authors of our current paper), our references contain 12.59\% woman(first)/woman(last), 22.68\% man/woman, 22.68\% woman/man, and 42.06\% man/man categorization. 
This method is limited in that a) names, pronouns, and social media profiles used to construct the databases may not, in every case, be indicative of gender identity and b) it cannot account for intersex, non-binary, or transgender people.
We look forward to future work that could help us to better understand how to support equitable practices in science. 

\section{Acknowledgements}
We thank A. Kahn, L. Parkes, M. Bertolero, P. Fotiadis, and S. Patankar for helpful discussions.
E.G.T and D.S.B. are supported by the National Science Foundation Materials Research Science and Engineering Center at University of Pennsylvania (NSF grant DMR-1120901), and the Paul G. Allen Family Foundation.
The views and conclusions contained in this document are solely those of the authors.

\bibliographystyle{apsrev4-2}
\bibliography{glassy,dsb_neuro,dsb_mat,misc,cieslak_adds,bassett_adds,diversity}

\end{document}